\documentclass[prl,aps,twocolumn,preprintnumbers,amsmath,amssymb,showpacs]{revtex4}
\usepackage{graphicx}\usepackage{epsfig}
\begin{document}
\title{
    Isoscaling as a measure of Symmetry Energy in the Lattice Gas Model
}
%
%
\author{G. Lehaut}
\author{F. Gulminelli 
}
\author{O. Lopez}
\affiliation{LPC Caen, ENSICAEN, Universit\'e de Caen, CNRS/IN2P3, Caen, France}
%
%
\begin{abstract}
	The energetic properties of nuclear clusters inside a low-density, finite-temperature medium are studied 
	with a Lattice Gas Model including isospin dependence and Coulomb forces. 
	Important deviations are observed respect to the Fisher approximation of an ideal gas of non-interacting 
	clusters, but the global energetics can still be approximately expressed in terms of a simple modified 
	energy-density functional. The multi-fragmentation regime appears dominated by combinatorial effects in this
	model, but the isoscaling of the largest fragment in low energy collisions 
	appears a promising observable for the experimental measurement of the symmetry 
	energy.
\end{abstract}
%
%
%
%
 \maketitle
%


In the different stages of the formation of a neutron star, from the explosion of the supernova 
to the cooling of the proto-neutron star, the baryonic matter is at finite temperature and densities
different from the normal equilibrium density of nuclei. 
To understand these phenomena, a considerable experimental effort is presently devoted 
to the determination of the nuclear energy-density fuctional as a function of 
temperature and density\cite{msu,texas,indra,high_energy}. Many investigations are concentrated 
on the isovector properties of the system, the so-called symmetry energy, which behavior 
is still largely unknown.
In the mean-field theory,
the symmetry-energy term can be approximately reproduced as a simple polynomial function of the density 
$c_{sym}(\rho)\propto \rho^\gamma$\cite{Chen};
%
%
finite temperature leads just to a fractional
occupation of single-particle levels, and as such does not modify the functional
behavior of the interaction energy.

The problem is that at low density 
the mean-field theory can be severely incorrect\cite{horowitz}.
In the heavy-ion collisions which are used to probe the density dependence of the symmetry energy,
the system under study is typically 
inhomogeneous; it is therefore not clear 
whether, even at thermal equilibrium, the associated energy functional only 
depends on the global density as obtained in a mean-field based picture.

A simple formula has also been proposed\cite{botvina} to extract directly 
the symmetry energy 
from measured cluster properties obtained
in the fragmentation of two systems 
of charge $Z_1,Z_2$, mass $A_1,A_2$ at the same temperature $T$:
\begin{equation}
4 \frac{c_{sym}}{T} = \frac{\alpha}{\left(Z_1^2/A_1^2\right)-\left(Z_2^2/A_2^2\right)}
\label{equBotvina} 
\end{equation}
where $\alpha$ is the so-called isoscaling parameter that can be 
measured from isotopic yields\cite{isoscaling}.

 However eq.(\ref{equBotvina}) has been derived in the framework of
 macroscopic statistical models~\cite{botvina}, where many-body correlations are
 supposed to be entirely exhausted by clusterisation, and it appears to be strongly affected by 
 conservation laws and combinatorial effects\cite{dasgupta,claudio}. 
 Moreover, the $c_{sym}$ coefficient appearing 
 in eq.(\ref{equBotvina}) should correspond to the symmetry free-energy\cite{natowitz}, 
 which is equivalent to the symmetry energy only in the $T\to 0$ limit.

To progress on these issues, it is interesting to consider a microscopic model 
simple enough to be exactly solvable through Monte-Carlo simulations without 
any mean-field or independent-cluster approximation.
In this paper, we study the temperature and density 
dependence of the symmetry energy coefficient 
and its connection to experimental 
observables in a Lattice Gas Model.
	
%


Let us consider a system composed of $N$ neutral and $Z$ charged
particles of mass $m=939 MeV$ occupying a cubic lattice of $V=8000$ cells with four degrees of
freedom : one discrete variable $\sigma_i$
for isospin ($\sigma_i=\pm 1$ for protons (neutrons), $\sigma_i=0$ if the site is unoccupied), 
and three continuous variables $\vec{p_i}$ for the momentum.
The Hamiltonian of the system follows:
\begin{equation}
H=\sum_{\langle i,j \rangle} \epsilon_{\sigma_i \sigma_j} \sigma_i \sigma_j
 + \sum_{\sigma_i= \sigma_j=1, 
 i \neq j} \frac{I_c}{r_{ij}} 
 + \sum_{i=1}^{L^3} \frac{p_i^2}{2m} \sigma_i^2
\label{H}
\end{equation}
where $<i,j>$ are nearest neighbor cells, $\epsilon_{\sigma_i \sigma_j}$ is the 
coupling between nearest neighbor 
($\epsilon_{11}=\epsilon_{-1-1}=0, \epsilon_{1-1}=5.5 MeV$), $I_c(=1.44MeV/fm)$ is 
the Coulomb coupling between protons, and $r_{ij}$ is the 
distance between sites $i$ and $j$.
The lattice spacing $r_0 = 1.8 fm$ has been chosen such that a full lattice 
occupation corresponds to the saturation density of symmetric nuclear matter 
$r_0^{-3}=\rho_0=0.17 fm^{-3}$.

This model has been already shown to be able to give a qualitative description
of nuclear fragmentation\cite{das}. Moreover the simplified version with only 
one type of uncharged particles is well-known to be isomorphous to the Ising model\cite{campi},
which makes LGM a paradigm of first and second order phase transitions in finite systems.

Calculations are made in the isobar canonical ensemble, which has been shown to
 be the correct canonical ensemble to describe unbound systems in the vacuum
~\cite{FGAnnalsPhys}.  
The partition function reads:
\begin{equation}
\Omega = \sum_{(n)} exp \left(- \beta \left( H^{(n)} + P  R^{3(n)} \right)  \right)
\label{partition}
\end{equation}
where the sum runs over all the possible realizations $(n)$ of the system, and 
$R^{3(n)}$ is the global extension of the system for each 
partition $(n)$ defined as:

\begin{equation}
R^{3(n)} = \frac{2 \left(\sum r_i^3 \sigma_i^2\right)^{(n)}}{\left(\sum{\sigma_i^2}\right)^{(n)}}
\label{equR}
\end{equation}
%
%
The partition sum (\ref{partition}) is numerically sampled for each given value of temperature
 and pressure with standard Metropolis techniques~\cite{Richert}.	

%

\begin{figure}[!htbp]
  \begin{center}	
     \includegraphics[width=1.\columnwidth]{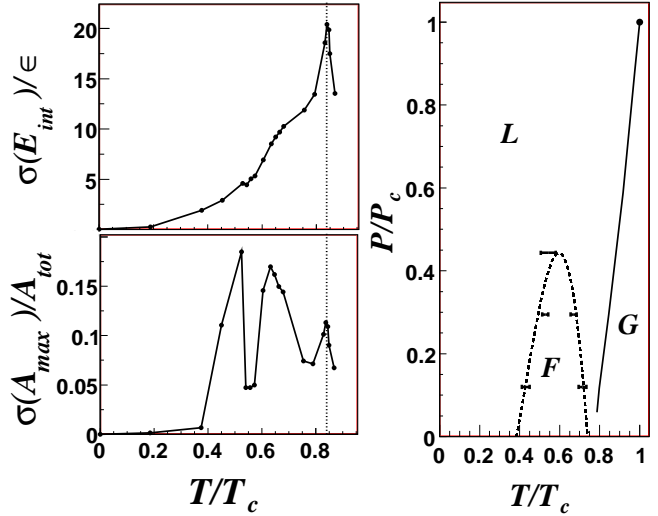}
  \end{center}
  \caption{Left side: fluctuation of the total interacting energy (upper part) and of the size of the largest cluster (lower part) 
  as a function of the temperature at $P=0.3P_c$, for a system with 75 neutrons and 54 protons. 
  Right side: phase diagram.
	The full line is the liquid (L) -gas (G) coexistence line and the point is the critical point.
	The symbols with horizontal error-bars give fission (F) -residue (L) transition points. The dashed line is to guide the eye.}
  \label{fig1}
\end{figure}

With only one type of particles, this model is well-known to exhibit a first order transition 
and a critical point ($T_c$,$P_c$), analogous to the liquid-gas transition.
The phase diagram of a finite system can be obtained within the isobar canonical ensemble
from the bimodality of the order parameter distribution~\cite{binder,noi}. At each pressure,
the total energy distribution, as well as the distribution of the size of the heaviest cluster produced in each event,
present two peaks of the same height at a temperature value,  
which is recognized as the transition
temperature~\cite{lehaut}. 
At this point the fluctuation are maximum, as shown in the left part of Fig.\ref{fig1} for a representative pressure. 
The ensemble of these transition points give the transition lines which are shown in the right part of Fig.\ref{fig1}.
We can see that adding a short-range isovector coupling and a long-range
repulsive interaction does not qualitatively modify the phenomenology 
of the liquid-gas transition (continuous line and full symbol). 
In particular both fluctuations peak at the same temperature (dotted line), showing that 
the fragmentation transition has a finite latent heat also for charged systems. 

But the phase diagram is considerably enriched respect to liquid-gas. 
Two extra transitions appear at lower
temperature which are specific to the nuclear phenomenology: inside the dashed curve, the system 
is splitted, without any energy jump, into two dominant fragments of similar size, 
which can be defined as hot fission.
This result is close to the findings of ref.\cite{chaudhuri}.
It is interesting to remark that bimodal distributions of the heaviest 
fragment have been recently observed experimentally\cite{pichon}.
%
In the following, we will only consider the systems above the residue-fission coexistence line.
	
%
%


The presence of phase transitions 
implies that 
Lattice Gas systems are strongly inhomogeneous and clusterized.
In this situation, it is not clear whether the energetics of the system
can be described by a macroscopic
parametrization depending only on the average density, as in the mean-field
approximation.

To explore this issue, 
we try a liquid-drop inspired macroscopic parametrization for the interaction energy 
of the system:
\begin{eqnarray}
  E_{int}^{LD}(\delta,\rho)&=& 
    \left ( a_v(\rho)+ c_{sym}^v(\rho)\delta^2\right ) A \nonumber \\
    &+& \left ( a_s(\rho)+ c_{sym}^s(\rho)\delta^2\right )A^{2/3} + \alpha_c(\rho)Z^2
  \label{LDParametrisation}
\end{eqnarray}
Here $\delta$ is the isospin asymmetry $\delta=(N-Z)/A$, $T$ is the temperature and
$\rho=A/(4/3\pi R^3)$ is an estimation of the average density of the system, 
where the mean cubic radius $R^3=\langle R^{3(n)} \rangle$ from eq.(\ref{equR}) is calculated excluding the 
monomers ($A=1$).

This liquid-drop parametrization (LD) uses five macroscopic 
density dependent parameters to be fitted to the model: $a_v$ is associated to 
the volume energy, $a_s$ corresponds to the surface energy, $c_{sym}^v$ and $c_{sym}^s$
give the bulk and surface part of the symmetry energy, and $\alpha_c$ corresponds to the 
Coulomb interaction.

\begin{figure}[!htbp]
  \begin{center}
    \includegraphics[width=1\columnwidth]{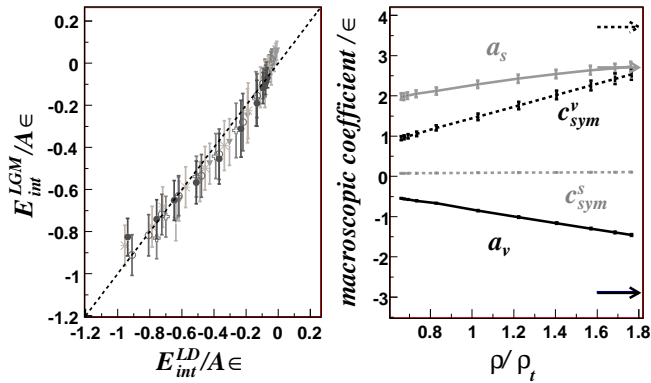}
  \end{center}
  \caption{Left part: Best fit obtained from the macroscopic parametrization eq.(\ref{LDParametrisation}) 
    versus the exact average energy at various average volumes and temperatures evaluated by Metropolis. 
    Right part: Evolution of the macroscopic coefficients of eq.(\ref{LDParametrisation}) with the average density.
	The arrows give the zero temperature values of the macroscopic coefficients.}
  \label{fig2}
\end{figure}

Eleven different systems of mass number $A=150$ and isospin ratio ranging from 
$\delta=-1/3$ to $\delta=1/3$ are simulated at the pressure $P/P_c=0.3$ and temperatures ranging 
from $T/T_t=0.88$ to $T/T_t=1.1$. This leads to an average volume variation
between $\rho/\rho_t=0.7$ and $\rho/\rho_t=1.8$. For each simulation, only one event out of $2\times10^4$
is kept to minimize auto-correlations, and averages are taken over $10^6$ events after 
a thermalization stage of typically $6\times10^4$ events.  
The average energy calculated from the Metropolis simulation 
is confronted to the best fit obtained from eq.(\ref{LDParametrisation}) 
on the upper part of Figure~\ref{fig2}.
With a global $\chi^2/N_{dof}=5$, we can consider eq.(\ref{LDParametrisation})
as a reasonable approximation to the exact energetics of the systems.
This is a non-trivial result, considering that the quantity $\rho$ 
entering eq.(\ref{LDParametrisation}) is never equal to the local density of the system,
which is strongly fluctuating.

The density evolution of the macroscopic coefficients extracted from the best fit  
is plotted on the bottom part of Figure~\ref{fig2}. The arrows  
give the values obtained when the same fitting procedure is applied to the systems
in their ground state\cite{das,lehaut}.
We observe a decrease of all parameters with decreasing density, but the effect on the bulk 
terms is more important than the effect on the surface terms. 
Surprisingly, the contribution of the surface term to the symmetry energy appears to be negligible. 
This is consistent with the findings of ref.\cite{ono}. 

This result is encouraging for the experimental effort of extracting the nuclear matter $c_{sym}$ out
of nuclear collisions: in the framework of this model, the low values extracted\cite{msu,texas,indra}
cannot be due to trivial surface effects\cite{danielewicz,raduta}.

The other interesting point is that the whole temperature dependence of the energetics is
entirely embedded in the density dependence of the macroscopic parameters exactly like in the mean-field theory\cite{baoan}
\footnote{The same would not be true if we were to calculate the average free energy $f=\langle e \rangle - T s$ 
instead than the average energy $\langle e \rangle$, which is needed as an input of macroscopic statistical models. This latter 
quantity would depend specifically on the temperature because of the entropy term.}.
This is another non-trivial result, because different partitioning of the system could be associated to the same average 
spatial extension at different temperatures. What Fig.~\ref{fig2} demonstrates is that 
, at least in the framework of classical physics and in the considered temperature-density range, 
the density functional approach can be a very good approach even in the presence of high-order correlations.  
%

This result suggests that it may be possible to extract 
the density dependence of the symmetry energy even if 
the system is clusterized.	

In order to explore this possibility, 
we now turn to examine the cluster distribution by looking at the 
isoscaling observable.
It is empirically well known that the ratio $R_{21}(N,Z)$ of isotope 
yields $Y_i(N,Z)$ measured in two reactions labeled ($1$,$2$) at the same incident energy but 
different in isospin has an exponential behavior \cite{isoscaling} according 
to :
\begin{equation}
R_{21} (N,Z) = \frac{Y_2(N,Z)}{Y_1(N,Z)}\propto exp \left( \alpha N + \beta Z \right)
\label{equIsoscaling}
\end{equation}
where $\alpha$ and $\beta$ are the isoscaling parameters.
It is reasonable to imagine that an observable like $\alpha$ which measures the fragment
isotopic content should be sensitive to the symmetry energy, and in particular to its 
density dependence\cite{msu}. Going a step forward, one may hope to extract 
$c_{sym}$ directly from the measured isoscaling through eq.(\ref{equBotvina}).

\begin{figure}[!htbp]
  \begin{center}
    \includegraphics[width=1\columnwidth]{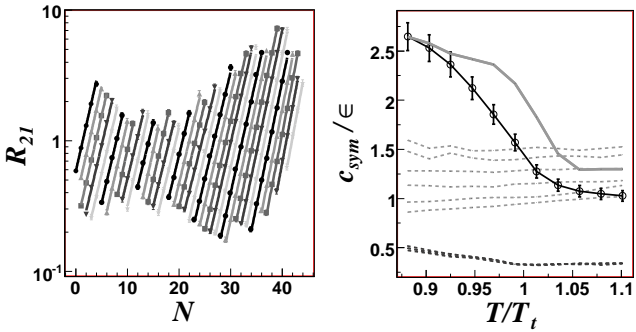}
  \end{center}
  \caption{Left part: isotopic ratio as a function of the neutron number at $T/T_t=0.95$ for the systems
  $(N=75,Z=75)$ and $(N=91,Z=59)$.
  Right part: 
  bold black line: symmetry energy of LGM.
  The other lines give different estimation of $c_{sym}$ from isoscaling.
  Bold grey line: symmetry energy from the
  analysis of the biggest fragment.
  Dashed grey lines: eq.(~\ref{equFrag}),
  dashed black line: eq.(~\ref{equBotvina}). }
\label{fig3}
\end{figure}

It has already been observed that isoscaling is well respected
in the LGM in the isochore ensemble~\cite{Ma,gupt3}.
The same is true 
also at constant pressure, 
as shown for a representative case in the 
left part of Fig.~\ref{fig3}.
The $\alpha$ parameter is almost constant with the fragment charge, and
decreases with increasing temperature, which corresponds to decreasing density
in the isobar ensemble.

The symmetry coefficient extracted from eq.(\ref{equBotvina}) using the value 
of $\alpha$ averaged from $Z=2$ to $Z=7$ 
is plotted on Fig.~\ref{fig3}. We can see that the resulting parameter is 
almost constant and completely disagrees with the symmetry energy of the model, as already observed in ref.\cite{gupt3}. 
An alternative approximate formula was derived in the fragmentation regime in refs.\cite{ono,raduta}:

\begin{equation}
4 \frac{c_{sym}(Z)}{T} = \frac{\alpha(Z)}{\left(Z^2/<A>_1^2\right)-\left(Z^2/<A>_2^2\right)}
\label{equFrag}
\end{equation}

where $Z$ is the charge of the considered fragment and $<A>_i$ 
is the mean mass of this fragment as obtained in system $i=1,2$.
In this expression $c_{sym}$ corresponds to the fragment symmetry energy, which may differ from 
the one of the source, for instance because of different surface effects\cite{danielewicz,raduta}.
Different values of $c_{sym}(Z)$ (dashed lines in Fig.\ref{fig3}) are observed for the different clusters charges 
applying eq.(\ref{equFrag}). Indeed $\alpha$ is almost independent of $Z$, while the isotopic content of the fragments
does depend on the fragment size: this is the well known fractionation phenomenon\cite{high_energy,raduta,plb}.
Since the resulting $c_{sym}(Z)$ are constant for the different 
densities, eq.(\ref{equFrag})
appears also inadequate to reproduce the symmetry energy of the model.
%
%
%
As discussed in ref.\cite{claudio}, the weak sensitivity of the $\alpha$ parameter to $c_{sym}$ 
is due to the fact that light cluster yields 
are dominated by combinatorial probabilities, and
do not reflect the thermodynamic properties of the system. 

The mass distribution of LGM is dominated by a large percolating cluster
which is the order parameter of the fragmentation transition, and contains most information on the thermodynamics\cite{big}.
One may then expect that the isotopic distribution of the heaviest cluster produced in each event may be more sensitive to 
the symmetry energy of the fragmenting system.

The result of applying eq.(\ref{equFrag}) to the largest cluster is plotted on Figure~\ref{fig3} as a continuous grey line.
We observe a much better agreement, except close to the transition temperature.
This may be due to the fact that very huge energy fluctuations are observed at the transition temperature in this canonical 
model (see Fig.\ref{fig1}), while the energy functional eq.(\ref{LDParametrisation}) depends only on average quantities. 
We expect that a better agreement will be obtained with a parametrization of the energy functional as a function of excitation
energy and density, instead than temperature and density.
This will further allow direct comparison with experimental data 
and will be the object of future investigations~\cite{lehaut}. 
%
%
%

To conclude, we have presented in this paper 
a study on the fragment properties at finite temperature and low density in the framework of a simple exactly 
solvable model 
.


We have shown that even in thermodynamic configurations close to a phase transition, where the system
is highly dishomogeneous and clusterized, the exact average energy can be well described 
as a simple functional of the overall average density of the system.
We have especially focussed our interest on the possible measurement of the density dependence of the symmetry 
energy, which is a topic of strong current interest in the nuclear physics 
community. 
We have shown that, in the framework of this model, the evolution of the symmetry energy term with 
the temperature and/or density can be traced with the help of the isotopic distribution of the 
largest cluster produced in each fragmentation event. 

Such measurements are presently undertaken by different experimental groups with the MARS recoil separator 
at Texas A\&M\cite{souliotis}
and the VAMOS spectrometer coupled with the INDRA $4-\pi$ array at GANIL~\cite{Chbihi}, and different experiments
in this line are planned with future RiB's facilities\cite{fazia}. Such data should allow an important advance
in the understanding of the functional behavior of the nuclear symmetry energy at finite temperature.

%
%

%
%
%
%
\end{document}